\documentclass[letterpaper,10pt]{article}

\usepackage{amsmath}
\usepackage{amssymb}
\usepackage{psfrag}
\usepackage[dvips]{graphicx}
\usepackage{epsfig}

\newcommand{\dd}{\textrm{d}}
\newcommand{\im}{{\mathbb{I}}{\mathrm{m}}}
\newcommand{\re}{{\mathbb{R}}{\mathrm{e}}}

\author{A.\ L\'opez-Ortega\thanks{alopezo@esfm.ipn.mx} \\ 
Departamento de F\'{\i}sica. Escuela Superior de F\'{\i}sica y Matem\'aticas. \\
Instituto Polit\'ecnico Nacional. \\
Unidad Profesional Adolfo L\'opez Mateos. Edificio 9. \\
M\'exico, D.\ F., M\'exico. \\
C.\ P.\ 07738 }

\title{Electromagnetic quasinormal modes of an asymptotically Lifshitz black hole}

\begin{document}

\maketitle

\begin{abstract}
 
Motivated by the recent interest in the study of the spacetimes that are asymptotically Lifshitz and in order to extend some previous results, we calculate exactly the quasinormal frequencies of the electromagnetic field in a $D$-dimensional asymptotically Lifshitz black hole. Based on the values obtained for the quasinormal frequencies we discuss the classical stability of the quasinormal modes. We also study whether the electromagnetic field possesses unstable modes in the $D$-dimensional Lifshitz spacetime. 



\end{abstract}

\section{Introduction}
\label{s: Introduction}

The oscillations of perturbation fields in gravitational systems have been studied for a long time. In particular for the black holes we know that a perturbation oscillates with damped oscillations called quasinormal modes (QNM) when we impose the appropriate conditions at the boundaries (the event horizon and the asymptotic region) \cite{Kokkotas:1999bd}--\cite{Konoplya:2011qq}. The study of the QNM is useful to analyze the classical stability of the black holes, to calculate relevant quantities in the AdS-CFT correspondence, and to test the proposals that determine the size of the entropy quantum for the event horizon from the asymptotic value of the quasinormal frequencies (QNF) \cite{Kokkotas:1999bd}--\cite{Konoplya:2011qq}.

Usually it is considered that the more relevant perturbation to be analyzed is the gravitational one, but we also study the QNM of test fields as the electromagnetic, Dirac, or Klein-Gordon perturbations. We believe that it is useful to study the QNM of these test fields since the analysis is easier than for the gravitational perturbations and the results can indicate the existence of possible issues.

It is known that in many condensed matter systems at the critical points (the Lifshitz fixed points) the space and time scale in the form 
\begin{equation} \label{e: anisotropic scale}
 t \to \lambda^{2z} t, \qquad x_i \to \lambda^2 x_i,
\end{equation} 
where $z >1$ is the dynamical critical exponent. To generalize the AdS-CFT correspondence to condensed matter systems with the anisotropic scale invariance (\ref{e: anisotropic scale}), recently there is interest in determining the properties of spacetimes with metrics that at large $r$ asymptote to the so called $D$-dimensional Lifshitz metric \cite{Kachru:2008yh}, \cite{Balasubramanian:2009rx}, \cite{Giacomini:2012hg}
\begin{equation} \label{e: Lifshitz metric}
 \dd s^2 = -\frac{r^{2 z}}{l^{2 z}} \dd t^2 + \frac{l^2}{r^2} \dd r^2 + r^2 \dd  \vec{x} \cdot \dd \vec{x} ,
\end{equation} 
where $l$ is a positive constant and $\dd  \vec{x} \cdot \dd \vec{x}$ is the line element of the $(D-2)$-dimensional Euclidean plane $\mathbb{R}^{D-2}$.  In the spacetime (\ref{e: Lifshitz metric}) the parameter $z$ coincides with the dynamical critical exponent of the condensed matter theory that we try to describe in a holographic way.

At present time  we know several black holes that asymptote to the spacetime (\ref{e: Lifshitz metric}). For $z=2$ a widely studied $D$-di\-men\-sion\-al asymptotically Lifshitz  black hole has a metric given by \cite{Balasubramanian:2009rx}, \cite{Giacomini:2012hg}
\begin{equation} \label{e: black hole Lifshitz}
\dd s^2 = -\frac{r^4}{l^4} \left( 1- \frac{r_+^2}{r^2} \right) \dd t^2 + \frac{l^2 \dd r^2}{r^2 -r_+^2} + r^2 \dd  \vec{x} \cdot \dd \vec{x}. \end{equation} 
This asymptotically Lifshitz black hole is a solution to the equations of motion for a Lagrangian with scalar and gauge  fields \cite{Balasubramanian:2009rx} or for a Lagrangian with higher-curvature terms \cite{Giacomini:2012hg}. In order to study the classical stability of the asymptotically Lifshitz black hole (\ref{e: black hole Lifshitz}) under small perturbations, Giacomini, et al.\ \cite{Giacomini:2012hg}  calculate exactly the QNF of a test Klein-Gordon field propagating in this background. Based on the values of the QNF and using a similar method to that proposed by Horowitz and Hubeny \cite{Horowitz:1999jd} for asymptotically anti-de Sitter black holes, they show that the QNM of the Klein-Gordon field are stable.  See Refs.\ \cite{Gonzalez:2012de}-\cite{Lepe:2012zf} for related works in which the QNF of asymptotically Lifshitz black holes are determined. We call attention to Refs.\ \cite{Becar:2007hu}--\cite{LopezOrtega:2007vu} where the QNF of higher dimensional backgrounds are calculated exactly. 

Furthermore in the study of gravity duals for non-relativistic condensed matter systems we expect that the QNM of the asymptotically Lifshitz black holes play a role similar to that of the QNM for asymptotically anti-de Sitter black holes in the analysis of the AdS-CFT correspondence \cite{Birmingham:2001pj}. Therefore we believe that it is a valuable exercise to calculate the QNF of the asymptotically Lifshitz spacetimes as the $D$-dimensional black hole (\ref{e: black hole Lifshitz}).

Based on the results by Kodama and Ishibashi \cite{Kodama:2003kk} for the coupled electromagnetic and gravitational perturbations of maximally symmetric spacetimes, in the $D$-dimensional Lifshitz black hole (\ref{e: black hole Lifshitz}) we simplify the equations of motion for a test electromagnetic field to a pair of radial differential equations. It is convenient to comment that, as in Ref.\ \cite{Giacomini:2012hg}, we consider the Lifshitz black hole (\ref{e: black hole Lifshitz}) as a solution to the equations of motion derived from a Lagrangian with higher curvature terms in order to consider the electromagnetic field in a consistent way as a test field. Using the pair of radial equations for the test electromagnetic field we extend the results by Giacomini, et al.\ \cite{Giacomini:2012hg} and calculate exactly its QNF when it propagates in the asymptotically Lifshitz black hole (\ref{e: black hole Lifshitz}). We think that this calculation is a step towards establishing its classical stability under small perturbations. 

We organize this paper as follows. In Sect.\ \ref{s: electromagnetic fields}, following Kodama and Ishibashi, we write the equations that satisfy the vector type and scalar type electromagnetic test fields in a $D$-dimensional maximally symmetric spacetime. In Sects.\ \ref{s: vector type QNF}, \ref{s: scalar type QNF}, and \ref{s: scalar n 3 4 5} we calculate exactly the QNF of the vector type and scalar type electromagnetic fields in the $D$-dimensional asymptotically Lifshitz black hole (\ref{e: black hole Lifshitz}). We discuss our main results in Sect.\ \ref{s: discussion}. In the Appendix we determine whether the test electromagnetic field possesses unstable modes in the $D$-dimensional Lifshitz spacetime (\ref{e: Lifshitz metric}).

\section{Electromagnetic fields}
\label{s: electromagnetic fields}

Consider a $D$-dimensional spacetime ($D=n+2$) with a line element of the form \cite{Kodama:2003kk}
\begin{equation} \label{e: line element general}
 \dd s^2 = g_{ab} (y) \dd y^a \dd y^b + r^2 (y) \dd \Omega_n^2 ,
\end{equation} 
where $ g_{ab} (y)$ is the metric of a bidimensional spacetime, $a,b=1,2,$ and $\dd \Omega_n^2 = \hat{\gamma}_{ij} (\hat{z}) \dd \hat{z}^i \dd \hat{z}^j $, $i,j=1,2,\dots,n$, is the line element of a $n$-dimensional maximally symmetric Einstein manifold whose Ricci tensor satisfies\footnote{We put a hat on the quantities defined on the base manifold $\dd \Omega_n^2$, for example, we denote the covariant derivative by $\hat{D}_j$.} $\hat{R}_{ij}=(n-1)\hat{K} \hat{\gamma}_{ij}$, where $\hat{K}$ is a constant related to its sectional curvature. In what follows we assume that $n \geq 2$ ($D \geq 4$).

The free Maxwell equations are the two equations
\begin{equation} \label{e: Maxwell equations}
 \dd \mathbb{F} = 0, \qquad \qquad \nabla_\nu F^{\mu \nu} = 0, 
\end{equation} 
where $\mathbb{F} = F_{\mu \nu} \dd x^\mu \wedge \dd x^\nu / 2$ is the electromagnetic field strength. As is well known, the first equation in (\ref{e: Maxwell equations}) implies that we can describe the electromagnetic field in terms of a vector potential, and therefore it decomposes into a vector type and a scalar type perturbation. From the results of Ref.\ \cite{Kodama:2003kk} we get that in the uncharged spacetime (\ref{e: line element general}) the Maxwell equations (\ref{e: Maxwell equations}) simplify to two decoupled equations when we decompose the electromagnetic field into different tensorial types on the $n$-dimensional manifold $\dd \Omega_n^2$.  

Following Kodama and Ishibashi we deduce that in the uncharged spacetime of the form (\ref{e: line element general}) the free Maxwell equations for the vector type electromagnetic field simplify to (see Eq.\ (4.23) of Ref.\ \cite{Kodama:2003kk})
\begin{equation} \label{e: vector type}
 \frac{1}{r^{n-2}} D_a (r^{n-2} D^a \mathbb{A}_V) - \frac{\hat{k}_V^2 + (n-1) \hat{K} }{r^2 }\mathbb{A}_V = 0 ,
\end{equation} 
where $D_a$ is the covariant derivative on the bidimensional spacetime with metric $g_{ab} (y)$ and the function $\mathbb{A}_V$ depends on the coordinates $y^a$ of the bidimensional manifold. If $\hat{\mathbb{V}}_i$ denotes the vector harmonics on the manifold $\dd \Omega_n^2$, that is,
\begin{equation} \label{e: vector harmonics}
 (\hat{D}^j  \hat{D}_j + \hat{k}_V^2 ) \hat{\mathbb{V}}_i = 0, \qquad \qquad \hat{D}^j \hat{\mathbb{V}}_j = 0 ,
\end{equation} 
then $\hat{k}_V^2$ are the eigenvalues of the vector harmonics on the base manifold $\dd \Omega_n^2$. From the definitions (\ref{e: vector harmonics}) it is possible to show that the vector harmonics satisfy \cite{Kodama:2003kk}
\begin{equation}
 \hat{D}^j ( \hat{D}_i \hat{\mathbb{V}}_j - \hat{D}_j  \hat{\mathbb{V}}_i) = (\hat{k}_V^2 + (n-1)\hat{K} ) \hat{\mathbb{V}}_i .
\end{equation} 

For the scalar type electromagnetic field propagating in an uncharged spacetime of the form (\ref{e: line element general}) the free Maxwell equations simplify to (see Eq.\ (5.20) of Ref.\ \cite{Kodama:2003kk})
\begin{equation}\label{e: scalar type}
 r^{n-2} D_a \left(\frac{ D^a \mathbb{A}_S }{r^{n-2}} \right) - \frac{\hat{k}^2}{r^2} \mathbb{A}_S  = 0,
\end{equation} 
where $\mathbb{A}_S$ is a function of the coordinates $y^a$, and $\hat{k}^2$ are the eigenvalues of the Laplacian on the $n$-dimensional manifold $\dd \Omega_n^2$, that is,
\begin{equation}
 (\hat{D}^j  \hat{D}_j + \hat{k}^2 ) \hat{\mathbb{S}} = 0 ,
\end{equation} 
with $\hat{\mathbb{S}}$ denoting the scalar harmonics on the base manifold $\dd \Omega_n^2$.

Since for the $D$-dimensional Lifshitz black hole (\ref{e: black hole Lifshitz}) we are assuming that the base manifold $\dd \Omega^2_n$ is $\mathbb{R}^n = \mathbb{R}^{D-2}$, then $\hat{K} = 0$ and the eigenvalues $\hat{k}_V$ and $\hat{k}$ have the properties: $\hat{k}_V^2$ ($\hat{k}^2$) is continuous with $\hat{k}_V^2 > 0 $ ($\hat{k}^2 > 0$) \cite{Kodama:2003kk}. In contrast to Kodama and Ishibashi \cite{Kodama:2003kk}, based on the form of the black hole metric (\ref{e: black hole Lifshitz}), we consider that the line element of the bidimensional spacetime in the formula (\ref{e: line element general}) takes the form
\begin{equation} \label{e: bidimensional metric}
 \dd s^2_{2} = g_{ab} \dd y^a \dd y^b= -F \dd t^2 + \frac{\dd r^2}{G},
\end{equation} 
with $F$ and $G$ functions of the coordinate $r$. 

Taking 
\begin{equation}
 \mathbb{A}_V = \frac{\Phi_V}{r^{n/2 -1}}, \qquad \qquad \mathbb{A}_S =  r^{n/2-1} \Phi_S ,
\end{equation} 
we find that Eq.\ (\ref{e: vector type}) for the vector type electromagnetic field simplifies to 
\begin{align} \label{e: vector type two-dimensional}
 D_a D^a  \Phi_V - \frac{n-2}{4r} \frac{\dd G}{\dd r}  \Phi_V - \frac{(n-2)(n-4)G}{4 r^2}  \Phi_V - \frac{n-2}{4r} \frac{G}{F} \frac{\dd F}{\dd r} \Phi_V - \frac{\hat{k}_V^2}{r^2} \Phi_V= 0,  
\end{align}
and Eq.\ (\ref{e: scalar type}) for the scalar type electromagnetic field becomes
\begin{align} \label{e: scalar type two-dimensional}
 D_a D^a \Phi_S - \frac{(n-2)n}{4} \frac{G}{r^2} \Phi_S + \frac{\dd G}{\dd r} \frac{n-2}{4r} \Phi_S + \frac{G}{r F} \frac{\dd F}{\dd r} \frac{n-2}{4} \Phi_S -\frac{\hat{k}^2}{r^2} \Phi_S = 0 .
\end{align}
Equations (\ref{e: vector type two-dimensional}) and (\ref{e: scalar type two-dimensional}) are the basis for the rest of our work.

A similar simplification of the Maxwell equations appears in Ref.\ \cite{Crispino:2000jx}, but we point out that Crispino, et al.\ only consider $D$-dimensional spherically symmetric spacetimes.

\section{Quasinormal frequencies of the vector type electromagnetic field}
\label{s: vector type QNF}

As in Ref.\ \cite{Giacomini:2012hg} we define the QNM of the electromagnetic field in the asymptotically Lifshitz the black hole (\ref{e: black hole Lifshitz}) as the modes satisfying the boundary conditions (see also \cite{Gonzalez:2012de}--\cite{Lepe:2012zf} and Sect.\ \ref{s: scalar n 3 4 5} below)
\begin{enumerate}
 \item[a)] They are purely ingoing near the horizon.
\item[b)] They go to zero at the asymptotic region.
\end{enumerate}

Considering that for the bidimensional metric (\ref{e: bidimensional metric}) we get
\begin{equation}
 D_a D^a f = - \frac{1}{F} \partial_t^2 f + \sqrt{\frac{G}{F}} \partial_r (\sqrt{F G } \partial_r ) f, 
\end{equation} 
and taking 
\begin{equation} \label{e: Phi vector}
 \Phi_V = \textrm{e}^{-i \omega t} R_V (r) ,
\end{equation} 
we find that Eq.\ (\ref{e: vector type two-dimensional}) simplifies to the ordinary differential equation for the radial function $R_V$
\begin{align} \label{e: radial vector type}
& r(r^2 - r_+^2) \frac{\dd }{\dd r} \left( r(r^2 - r_+^2) \frac{\dd R_V}{\dd r} \right) +(\omega l^3)^2 R_V - \left( \frac{n-2}{2}((r^2-r_+^2)^2 + r^2(r^2-r_+^2))  \right.  \\
& \left. + \frac{(n-2)(n-4)}{4}(r^2-r_+^2)^2 + \frac{n-2}{2} r^2 (r^2 - r_+^2) +  \hat{k}_V^2l^2 (r^2 - r_+^2) \right) R_V = 0 . \nonumber
\end{align}
As in Ref.\ \cite{Giacomini:2012hg}, to solve the previous differential equation we make the change of variable
\begin{equation} \label{e: change of variable}
 y = 1 - \frac{r_+^2}{r^2},
\end{equation} 
to obtain that the radial function $R_V$ must be a solution of the differential equation
\begin{align}
 \frac{\dd^2 R_V}{\dd y^2} + \frac{1}{y} \frac{\dd R_V}{\dd y} + \left( \frac{\tilde{\omega}^2}{y^2} - \frac{n^2-4}{16} \frac{1}{(1-y)^2} -\left( \tilde{k}_V^2 + \frac{n-2}{4} \right) \frac{1}{y(1-y)}  \right)R_V=0 ,
\end{align}
where 
\begin{equation}
 \tilde{\omega} = \frac{\omega l^3}{2 r_+^2}, \qquad \qquad \tilde{k}_V = \frac{\hat{k}_V l}{2 r_+}.
\end{equation} 
Notice that for $r > r_+$ we obtain that $y \in (0,1)$.

Proposing that the radial function $R_V$ takes the form 
\begin{equation}
 R_V = (1-y)^{B_V} y^{A_V} \tilde{R}_V,
\end{equation} 
we find that if the parameters $B_V$ and $A_V$ are solutions of the equations
\begin{equation}
 B_V^2 - B_V - \frac{n^2-4}{16}=0, \qquad \qquad A_V^2 + \tilde{\omega}^2 = 0,
\end{equation} 
then the function $\tilde{R}_V$ must be a solution of the differential equation 
\begin{align}
 y (1-y)\frac{\dd^2 \tilde{R}_V}{\dd y^2} +& (2 A_V +1 - (2 A_V + 2 B_V + 1)y )\frac{\dd \tilde{R}_V}{\dd y} \\
&- \left(\tilde{k}_V^2 + \frac{n-2}{4} + B_V + 2 A_V  B_V \right) \tilde{R}_V = 0, \nonumber
\end{align}
which is a hypergeometric differential equation 
\begin{equation} \label{e: hypergeometric differential}
y(1-y) \frac{{\rm d}^2 \tilde{R}_V}{{\rm d} y^2} + (c_V - (a_V +b_V + 1)y)\frac{{\rm d} \tilde{R}_V}{{\rm d} y} - a_V b_V \tilde{R}_V   = 0,
\end{equation}
with the parameters $a_V$, $b_V$, and $c_V$ given by
\begin{align} \label{e: a b c vector type}
 &a_V = A_V + B_V + \frac{1}{2} \sqrt{-4 \tilde{\omega}^2 + \frac{(n-2)^2}{4} - 4 \tilde{k}_V^2} , \qquad  c_V = 2 A_V + 1 , \nonumber \\
&b_V = A_V + B_V - \frac{1}{2} \sqrt{-4 \tilde{\omega}^2 + \frac{(n-2)^2}{4} - 4 \tilde{k}_V^2} . 
\end{align} 

In what follows we take the parameters $A_V$ and $B_V$ as
\begin{equation}
 B_V= \frac{1}{2} + \frac{n}{4}, \qquad \qquad   A_V = i \tilde{\omega}.
\end{equation} 
Therefore, assuming that the quantity $c_V$ is not an integer, we see that the radial function $R_V$ takes the form \cite{Abramowitz-book}, \cite{Guo-book}
\begin{align} \label{e: radial function vector type}
 R_V = (1-y)^{1/2 + n/4} y^{i \tilde{\omega}} ( &C_1 \,\,  {}_{2}F_{1}(a_V,b_V;c_V;y) \nonumber \\
&+ C_2 y^{1-c_V} {}_{2}F_{1}(a_V-c_V+1,b_V-c_V+1;2-c_V;y) ) ,
\end{align} 
where ${}_{2}F_{1}(a_V,b_V;c_V;y)$ denotes the hypergeometric function, and $C_1$, $C_2$ are constants.\footnote{ We must notice that in the distinct formulas where they appear, the values of the constants $C_1$ and $C_2$ may be different.}

To calculate the QNF we first impose the boundary condition a) near the horizon. We notice that the horizon is located at $y=0$, hence we find that near the horizon the radial function $R_V$ behaves as
\begin{equation} \label{e: radial vector near horizon}
 R_V \approx C_1 y^{i \tilde{\omega}} + C_2 y^{- i \tilde{\omega}} \approx C_1 \textrm{e}^{i \omega r_*} + C_2 \textrm{e}^{- i \omega r_*} ,
\end{equation} 
where $r_*$ denotes the tortoise coordinate and for the Lifshitz black hole (\ref{e: black hole Lifshitz}) it is equal to
\begin{equation}
 r_* = \frac{l^3}{2 r_+^2} \ln \left( 1 - \frac{r_+^2}{r^2} \right) =  \frac{l^3}{2 r_+^2} \ln (y)  .
\end{equation} 
Note that $r_* \in (-\infty, 0)$ for $r > r_+$.

As we choose a harmonic time dependence $\exp(-i \omega t)$ we get that in the expression (\ref{e: radial vector near horizon}) the term proportional to $C_1$ is an outgoing wave near the horizon and the term proportional to $C_2$ is an ingoing wave. Therefore to satisfy the boundary condition a) of the QNM we must take $C_1=0$, and the radial function $R_V$ simplifies to
\begin{align} \label{e: radial vector boundary horizon}
 R_V &= C_2  y^{- i \tilde{\omega}} (1-y)^{1/2 + n/4}  {}_{2}F_{1}(a_V-c_V+1,b_V-c_V+1;2-c_V;y) \nonumber \\
  & = C_2  y^{- i \tilde{\omega}} (1-y)^{1/2 + n/4} {}_{2}F_{1}(\alpha_V,\beta_V;\gamma_V;y),
\end{align}
that is, we define the quantities $\alpha_V$, $\beta_V$, and $\gamma_V$ by
\begin{equation} \label{e: alpha beta gamma vector}
 \alpha_V = a_V-c_V+1, \qquad \beta_V= b_V-c_V+1, \qquad \gamma_V = 2-c_V.
\end{equation} 
To impose the boundary condition b) of the QNM at the asymptotic region we must study the behavior of the radial function (\ref{e: radial vector boundary horizon}) near $y=1$. It is convenient to note that the parameters  $\alpha_V$, $\beta_V$, and $\gamma_V$ satisfy 
\begin{equation}
 \gamma_V - \alpha_V - \beta_V = - \frac{n}{2}, 
\end{equation} 
that is, $\gamma_V - \alpha_V - \beta_V$ is a negative integer for $n=2,4,6,\dots$, and a negative half-integer for $n=3,5,7,\dots$.

To study the behavior of the radial function near $y=1$ for $\gamma - \alpha - \beta$ different from an integer we exploit Kummer's formula \cite{Abramowitz-book}, \cite{Guo-book}
\begin{align} \label{e: Kummer property y 1-y}
{}_2F_1(\alpha,\beta;\gamma;y) &= \frac{\Gamma(\gamma) \Gamma(\gamma-\alpha-\beta)}{\Gamma(\gamma-\alpha) \Gamma(\gamma - \beta)} {}_2 F_1 (\alpha,\beta;\alpha+\beta+1-\gamma;1-y)  \\
&+ \frac{\Gamma(\gamma) \Gamma( \alpha + \beta - \gamma)}{\Gamma(\alpha) \Gamma(\beta)} (1-y)^{\gamma-\alpha -\beta} {}_2F_1(\gamma-\alpha, \gamma-\beta; \gamma + 1 -\alpha-\beta; 1 -y), \nonumber
\end{align}
whereas for $\gamma - \alpha - \beta = -m$, with $m$ a nonnegative integer, we must use \cite{Abramowitz-book}, \cite{Guo-book}
\begin{align} \label{e: hypergeometric property integer}
{}_2F_1 (\alpha,\beta;\gamma;y) &=   \frac{\Gamma(\gamma) \Gamma(m)}{\Gamma(\alpha)\Gamma(\beta)} (1-y)^{-m} \sum_{q=0}^{m-1}\frac{(\alpha-m)_q (\beta-m)_q}{q! (1-m)_q}(1-y)^q \\ 
&  + \frac{(-1)^{m+1} \Gamma(\gamma)}{\Gamma(\alpha-m)\Gamma(\beta-m)} \sum_{q=0}^\infty \frac{(\alpha)_q (\beta)_q}{q!(m+q)!}(1-y)^q  \nonumber \\
&\times [\ln(1-y) -\psi(q+1) -\psi(q+m+1)  +\psi(\alpha+q)+\psi(\beta+q)],  \nonumber
\end{align}
where $\psi(y)={\rm d} \ln \Gamma(y)/{\rm d}y$, $(\alpha)_0=1$, $(\alpha)_q=\alpha(\alpha+1)\cdots(\alpha+q-1)$ for $q \geq 1$, and for $m=0$ the term containing the sum with a finite number of terms does not appear. We must notice that this property of the hypergeometric function is valid for $\alpha, \beta \neq 0, -1, -2, \dots$, \cite{Guo-book}.

For $n=3,5,\dots$, the quantity $\gamma_V - \alpha_V - \beta_V$ is a half-integer, thus, using Kummer's formula (\ref{e: Kummer property y 1-y}) we obtain that near $y=1$ the radial function $R_V$ behaves as
\begin{align}
 R_V \approx (1-y)^{1/2 +n/4}  \frac{\Gamma(\gamma_V) \Gamma(\gamma_V-\alpha_V-\beta_V)}{\Gamma(\gamma_V-\alpha_V) \Gamma(\gamma_V - \beta_V)} + (1-y)^{1/2 -n/4}  \frac{\Gamma(\gamma_V) \Gamma(\alpha_V+\beta_V-\gamma_V )}{\Gamma(\alpha_V) \Gamma(\beta_V)} .
\end{align}
From this expression we point out that, as $y \to 1$, the first term fulfills the boundary condition b), but the second term diverges in this limit. Therefore to satisfy this boundary condition we must impose the condition
\begin{equation} \label{e: conditions vector type}
 \alpha_V = a_V-c_V+1 = - p, \qquad \textrm{or} \qquad \beta_V = b_V-c_V+1= -p, \quad \quad p=0,1,2,3,\dots 
\end{equation} 
From these equations we get that for $n=3,5,\dots$, the QNF of the vector type electromagnetic field are equal to 
\begin{equation} \label{e: QNF vector type}
 \omega_V = -i\frac{r_+^2}{l^3} \frac{1}{p+\tfrac{1}{2} + \tfrac{n}{4}} \left[ (p+\tfrac{1}{2})^2 + \frac{n}{2} (p+\tfrac{1}{2}) + \frac{n-1}{4} + \tilde{k}_V^2\right].
\end{equation}  
Notice that these QNF are purely imaginary, as the QNF of the Klein-Gordon field found in Ref.\ \cite{Giacomini:2012hg}.

For $n=2,4,6,\dots$, we define $n=2N$, with $N$ a positive integer ($N \geq 1$).\footnote{Notice that the parameter $N$ can take different values in the Appendix and in Sects.\ \ref{s: vector type QNF}, \ref{s: scalar type QNF}.} Hence we find that
\begin{equation}
 \gamma_V - \alpha_V - \beta_V = - N, 
\end{equation} 
and therefore, for $\alpha_V, \beta_V \neq 0, -1, -2, \dots$, from the property (\ref{e: hypergeometric property integer}), we get that at the asymptotic region the radial function (\ref{e: radial vector boundary horizon}) behaves as
\begin{equation} \label{e: radial vector asymptotic even}
 R_V \approx  (1-y)^{1/2 - N/2}  \frac{\Gamma(\gamma_V) \Gamma(N)}{\Gamma(\alpha_V)\Gamma(\beta_V)} \sum_{q=0}^{N-1}\frac{(\alpha_V-N)_q (\beta_V-N)_q}{q! (1-N)_q}(1-y)^q,
\end{equation} 
since the other term goes to zero as $y \to 1$. Thus to  impose the boundary condition b) we must satisfy Eqs.\ (\ref{e: conditions vector type}), but it contradicts one assumption of the property (\ref{e: hypergeometric property integer}). Hence using this property of the hypergeometric function to transform the radial function (\ref{e: radial vector boundary horizon}) we can not fulfill the boundary condition b) of the QNM.

The remaining option is to consider that the hypergeometric function of the radial function (\ref{e: radial vector boundary horizon}) is a polynomial, that is, $\alpha_V = -p$, (or $\beta_V=-p$), with $p=0,1,2,3,\dots$, \cite{Guo-book}, \cite{NIST-book}. Owing to the symmetry in the parameters $\alpha_V$ and $\beta_V$ of the hypergeometric function ${}_{2}F_{1}(\alpha_V,\beta_V;\gamma_V;y)$ we consider only $\alpha_V = -p$, (we shall obtain similar results for $\beta_V = -p$). Thus we assume that the radial function (\ref{e: radial vector boundary horizon}) takes the form 
\begin{equation} \label{e: radial function vector integer}
 R_V = C_2 y^{-i \tilde{\omega}} (1-y)^{1/2+n/4} {}_2F_1(-p,\beta_V;\gamma_V;y),
\end{equation} 
and we analyze whether it satisfies the boundary condition b) of the QNM. From the expressions (\ref{e: a b c vector type}) and (\ref{e: alpha beta gamma vector}) we notice that if $\alpha_V$ is an non-positive integer, then $\beta_V$ and $\gamma_V$ are not integers. 

Taking into account that for $p$ a nonnegative integer the hypergeometric function fulfills \cite{Guo-book}, \cite{NIST-book}
\begin{equation} \label{e: hypergeometric polynomial}
 {}_2F_1(-p,\beta;\gamma;z) = \frac{(\gamma - \beta)_p}{(\gamma)_p} {}_2F_1(-p,\beta;\beta-\gamma-p+1;1-z),
\end{equation} 
and that $(-\gamma)_p$ is equal to 
\begin{equation} \label{e: property symbol}
 (-1)^p (-\gamma)_p  = (\gamma - p + 1)_p ,
\end{equation} 
we get that the radial function (\ref{e: radial function vector integer}) transforms into
\begin{equation} \label{e: radial vector integer}
 R_V = C_2 \frac{(N+1)_p}{(-1)^p (\gamma_V)_p} y^{-i \tilde{\omega}} (1-y)^{1/2+n/4} {}_2F_1(-p,\beta_V;\beta_V-\gamma_V-p+1;1-y) .
\end{equation} 
Therefore near $y=1$ the radial function (\ref{e: radial vector integer}) behaves as
\begin{equation}
 R_V \approx (1-y)^{1/2 + n/4} ,
\end{equation} 
that fulfills the boundary condition b) of the QNM. Thus for $\alpha_V = -p$ (or $\beta_V = -p$) the radial function (\ref{e: radial vector integer}) satisfies the boundary condition b) and hence for $n=2,4,6,\dots,$ we also get the QNF (\ref{e: QNF vector type}) for the vector type electromagnetic field propagating in the $D$-dimensional Lifshitz black hole (\ref{e: black hole Lifshitz}).

Finally we note that for the QNF (\ref{e: QNF vector type}) of the vector type electromagnetic field the quantity $c_V$ is not an integer, as we assumed previously.

\section{Quasinormal frequencies of the scalar type electromagnetic field}
\label{s: scalar type QNF}

Here we calculate the QNF of the scalar type electromagnetic field propagating in the asymptotically Lifshitz black hole (\ref{e: black hole Lifshitz}). Taking 
\begin{equation} \label{e: Phi scalar}
 \Phi_S = \textrm{e}^{-i \omega t} R_S (r) ,
\end{equation} 
we find that Eq.\ (\ref{e: scalar type two-dimensional}) reduces to the ordinary differential equation for the radial function $R_S$
\begin{align} \label{e: radial scalar type}
 r(r^2 - r_+^2) \frac{\dd }{\dd r} &\left( r(r^2 - r_+^2) \frac{\dd R_S}{\dd r} \right) +(\omega l^3)^2 R_S + \left( \frac{n-2}{2}( (r^2-r_+^2)^2 + r^2(r^2-r_+^2))  \right.  \\
& \left. - \frac{(n-2)n}{4}(r^2-r_+^2)^2 + \frac{n-2}{2} r^2 (r^2 - r_+^2) - \hat{k}^2 l^2 (r^2 - r_+^2) \right) R_S = 0 . \nonumber
\end{align}

Making the change of variable (\ref{e: change of variable}) and taking the radial function $R_S$ as 
\begin{equation}
 R_S = (1-y)^{B_S} y^{A_S} \tilde{R}_S,
\end{equation} 
where the parameters $B_S$ and $A_S$ must be solutions to the equations
\begin{equation}
 B_S^2 - B_S - \frac{(n-2)(n-6)}{16}=0, \qquad \qquad A_S^2 + \tilde{\omega}^2 = 0,
\end{equation} 
we find that the function $\tilde{R}_S$ is a solution of the hypergeometric type differential equation (\ref{e: hypergeometric differential}) with parameters 
\begin{align} \label{e: a b c scalar type}
 &a_S = A_S + B_S + \frac{1}{2} \sqrt{-4\tilde{\omega}^2 + \frac{(n-2)^2}{4} - 4 \tilde{k}_S^2} , \qquad  c_S = 2 A_S + 1, \nonumber \\
&b_S = A_S + B_S - \frac{1}{2} \sqrt{-4\tilde{\omega}^2 + \frac{(n-2)^2}{4} - 4 \tilde{k}_S^2} ,  
\end{align} 
where $\tilde{k}_S = \hat{k} l /(2 r_+)$. In what follows we take 
\begin{equation}
B_S= \frac{1}{2} + \frac{|n-4|}{4} , \qquad \qquad   A_S = i \tilde{\omega}.
\end{equation} 
Therefore, assuming that the parameter $c_S$ is not an integer, we see that the radial function $R_S$ is equal to
\begin{align} \label{e: radial function scalar type}
 R_S = (1-y)^{1/2 + |n-4|/4} y^{i \tilde{\omega}} ( &C_1 \,\,  {}_{2}F_{1}(a_S,b_S;c_S;y) \nonumber \\
&+ C_2 y^{1-c_S} {}_{2}F_{1}(a_S-c_S+1,b_S-c_S+1;2-c_S;y) ) ,
\end{align} 
with $C_1$ and $C_2$ constants.

As for the vector type electromagnetic field, to obtain a purely ingoing field near the horizon we must take $C_1 = 0$ in the expression (\ref{e: radial function scalar type}) and hence the radial function $R_S$ that satisfies the boundary condition a) near the horizon is 
\begin{align} \label{e: radial scalar boundary horizon}
 R_S = C_2  y^{- i \tilde{\omega}} (1-y)^{1/2 + |n-4|/4} {}_{2}F_{1}(\alpha_S,\beta_S;\gamma_S;y),
\end{align}
with $\alpha_S = a_S-c_S+1$, $\beta_S = b_S-c_S+1$, and $\gamma_S = 2 - c_S$. We notice that the parameters $\alpha_S$, $\beta_S$, and $\gamma_S$ fulfill
\begin{equation} \label{e: gamma beta alpha scalar difference}
 \gamma_S - \alpha_S - \beta_S = - \frac{|n-4|}{2}, 
\end{equation} 
that is, it is a negative half-integer for $n=3,5,7,\dots$, and a non-positive integer for $n=2,4,6,\dots$

We first study the odd values of $n$ for which $\gamma_S - \alpha_S - \beta_S$ is a negative half-integer and we can use the Kummer formula (\ref{e: Kummer property y 1-y}) to get
\begin{align} \label{e: Kummer property scalar radial}
&R_S = C_2 y^{- i \tilde{\omega} } \left[ (1-y)^{1/2 + |n-4|/4}  \frac{\Gamma(\gamma_S) \Gamma(\gamma_S-\alpha_S-\beta_S)}{\Gamma(\gamma_S-\alpha_S) \Gamma(\gamma_S - \beta_S)} {}_2 F_1 (\alpha_S,\beta_S;\alpha_S+\beta_S+1-\gamma_S;1-y)  \right. \nonumber \\
&+ \left. \frac{\Gamma(\gamma_S) \Gamma( \alpha_S + \beta_S - \gamma_S)}{\Gamma(\alpha_S) \Gamma(\beta_S)} (1-y)^{1/2 - |n-4|/4} {}_2F_1(\gamma_S-\alpha_S, \gamma_S-\beta_S; \gamma_S + 1 -\alpha_S-\beta_S; 1 -y) \right].
\end{align}
Since we are considering $n$ odd, with $n \geq 3$, we see that the term proportional to $(1-y)^{1/2 + |n-4|/4}$ goes to zero as $y \to 1$, thus, this term satisfies the boundary condition b) of the QNM. For the other term of the formula (\ref{e: Kummer property scalar radial}) we note the following facts. For $n=3,5,$ the factor 
\begin{equation} \label{e: factor scalar type}
 (1-y)^{1/2 - |n-4|/4}
\end{equation} 
goes to zero as $y \to 1$. Thus for these two values of $n$ the purely ingoing radial function $R_S$ of the formula (\ref{e: radial scalar boundary horizon}) satisfies automatically the boundary condition b) at the asymptotic region. Hence for $n=3,5,$ we may conclude that for any value of the frequency $\omega$ we can fulfill the boundary conditions of the QNF, but this result is not easily acceptable, since it implies the existence of unstable QNM. We think that it is necessary a more careful study and analyze in more detail the behavior of the radial functions for these two values of the spacetime dimension. (See Sect.\ \ref{s: scalar n 3 4 5} below.)

For $n=7,9,\dots$, we get that the factor (\ref{e: factor scalar type}) diverges as $y \to 1$, and therefore to satisfy the boundary condition b) we must impose the condition
\begin{equation} \label{e: conditions scalar type}
 \alpha_S = a_S-c_S+1 = - p, \qquad \textrm{or} \qquad \beta_S = b_S-c_S+1= -p,  \qquad p=0,1,2,3,\dots ,
\end{equation} 
from which we get that the QNF of the scalar type electromagnetic field are equal to
\begin{equation} \label{e: QNF scalar type}
 \omega_S = -i\frac{r_+^2}{l^3} \frac{1}{p+ \tfrac{n-2}{4}} \left[ p^2 + \frac{n-2}{2} p  + \tilde{k}_S^2 \right] .
\end{equation}  

Now we consider $n$ even. From the expression (\ref{e: gamma beta alpha scalar difference}) we obtain that for $n=2$, $\gamma_S - \alpha_S - \beta_S= -1$, and for $n=4,6,\dots$, we get that $\gamma_S - \alpha_S - \beta_S= -(n/2 - 2)= -(N -2)$, with $n = 2 N$ as previously, but now $N$ is a positive integer satisfying $N \geq 2$. Notice that for $n=4$ we find that $\gamma_S - \alpha_S - \beta_S=0$. For  $\gamma_S - \alpha_S - \beta_S \neq 0$ ($n \neq 4$), if we exploit the property (\ref{e: hypergeometric property integer}) of the hypergeometric function to transform the radial function (\ref{e: radial scalar boundary horizon}), then we can not fulfill the boundary condition b) of the QNM. As for the vector type electromagnetic field, by taking $\alpha_S = -p$, (or $\beta_S = -p$), $p=0,1,2,\dots$, we seek to impose the boundary condition b) of the QNM. Thus we assume that the radial function (\ref{e: radial scalar boundary horizon}) takes the form 
\begin{equation} \label{e: scalar type integer even}
 R_S = C_2 y^{-i \tilde{\omega}} (1-y)^{1/2+|n-4|/4} {}_{2}F_{1}(-p,\beta_S;\gamma_S;y) .
\end{equation} 
From the formulas (\ref{e: a b c scalar type}) and (\ref{e: radial scalar boundary horizon}), we infer that if $\alpha_S$ is an integer, then $\beta_S$ and $\gamma_S$ are not integers. Considering the properties (\ref{e: hypergeometric polynomial}) and (\ref{e: property symbol}) we obtain that for $n$ even, $n \neq 4$, at the asymptotic region the radial function $R_S$ of the formula (\ref{e: scalar type integer even}) behaves as 
\begin{equation} \label{e: scalar type n even asymptotic}
 R_S \approx (1-y)^{1/2 + |n-4|/4} ,
\end{equation} 
that satisfies the boundary condition b) of the QNM. As we impose the condition $\alpha_S = -p$ (or $\beta_S=-p$) to get the behavior (\ref{e: scalar type n even asymptotic}), for $n=6,8,\dots$, we also get the QNF (\ref{e: QNF scalar type}) for the scalar type electromagnetic field.

To obtain the expression (\ref{e: QNF scalar type}) for the QNF we assume that $|n-4|=n-4$, but it is not true for $n=2$. Taking into account that for $n=2$, $|n-4|=-(n-4)$, from the conditions (\ref{e: conditions scalar type}) we get that for $n=2$ the QNF of the scalar type electromagnetic field are equal to (see also Sect.\ \ref{s: scalar n 3 4 5} below)
\begin{equation} \label{e: QNF scalar type n=2}
 \omega_S = -i\frac{r_+^2}{l^3} \frac{1}{p+ 1} \left[ p^2 + 2 p +1 + \tilde{k}_S^2 \right]. 
\end{equation} 
Notice that for the scalar type electromagnetic field the QNF (\ref{e: QNF scalar type}) and (\ref{e: QNF scalar type n=2})  are purely imaginary as the QNF (\ref{e: QNF vector type}) of the vector type electromagnetic field and those given in Ref.\ \cite{Giacomini:2012hg} for the Klein-Gordon field.

For $n=4$ we get that $\gamma_S - \alpha_S - \beta_S= 0$ and from the property (\ref{e: hypergeometric property integer}) with $m=0$  we find that the radial function $R_S$ takes the form 
\begin{align}
 R_S = C_2 y^{- i \tilde{\omega}} (-1)& \frac{\Gamma(\gamma_S)}{\Gamma(\alpha_S) \Gamma(\beta_S)} \sum_{q=0}^{\infty} \frac{(\alpha_S)_q (\beta_S)_q }{(q!)^2} (1-y)^{1/2+q} \nonumber \\ 
& \times \left( \psi(\alpha_S +q) + \psi(\beta_S+q)-2\psi(1+q)+\ln(1-y) \right),
\end{align}
that goes to zero as $y \to 1$, that is, this radial function fulfills the boundary condition b) of the QNM without additional conditions on the frequency. Therefore, in a similar way to $n=3, 5$, for $n=4$ we must make a more careful analysis of the radial function behavior as $r \to \infty$.

Finally, in a straightforward way we can verify  that for the QNF (\ref{e: QNF scalar type}) and (\ref{e: QNF scalar type n=2}) of the scalar type electromagnetic field the quantity $c_S$ is not an integer, as we assumed previously.

\section{Quasinormal frequencies for $n=3,4,5$}
\label{s: scalar n 3 4 5}

Up to this point we follow closely to Refs.\ \cite{Giacomini:2012hg}, \cite{Gonzalez:2012de}-\cite{Lepe:2012zf}, and impose the boundary conditions  a) and b) to determine the QNF of the asymptotically Lifshitz black hole (\ref{e: black hole Lifshitz}). Nevertheless the possibility of obtaining a continuous QNF  spectrum for the scalar type electromagnetic field in $n=3,4,5,$ as we comment in the previous section, calls us for a more careful analysis of the boundary conditions that have been imposed. 

It is well known that in asymptotically anti-de Sitter black holes we can impose different boundary conditions at the asymptotic region, (see for example Refs.\ \cite{Berti:2009kk}, \cite{Konoplya:2011qq}, \cite{Breitenlohner:1982jf}--\cite{Dias:2013sdc}). Thus in a similar way to the asymptotically anti-de Sitter black holes we study in detail the boundary conditions that we can impose on the electromagnetic field at the asymptotic region of the $D$-dimensional Lifshitz black hole (\ref{e: black hole Lifshitz}).

Although for the vector type electromagnetic field we get well defined QNF in the Lifshitz black hole (\ref{e: black hole Lifshitz}) with $n \geq 2$, we begin analyzing this perturbation type. From Eq.\ (\ref{e: radial vector type}) we obtain that as $r \to \infty$ the radial function $R_V$ is a solution to
\begin{equation} \label{e: radial vector asymptotic}
 \frac{\dd^2 R_V}{\dd r^2} + \frac{3}{r} \frac{\dd R_V}{\dd r} - \frac{n^2-4}{4} \frac{1}{r^2} R_V = 0,
\end{equation} 
that is a Euler type differential equation. Hence as $r \to \infty$ the radial function behaves as
\begin{equation} \label{e: vector asymptotic behavior}
 R_V \approx C_1 r^{n/2 -1} + C_2 r^{-(n/2+1)},
\end{equation} 
where $C_1$ and $C_2$ are constants (as before). For $n > 2$ we find that the first term of the formula (\ref{e: vector asymptotic behavior}) diverges as $r \to \infty$ and the second term goes to zero in the same limit. For these values of $n$ it is natural that we impose the boundary conditions a) and b) of Sect.\ \ref{s: vector type QNF}. In the same limit, from the expression (\ref{e: vector asymptotic behavior}), for $n=2$  we obtain that $R_V$ behaves as
\begin{equation} \label{e: n=2 general asymptotic}
 R_V \approx C_1 + \frac{C_2}{r^2},
\end{equation} 
that is, both solutions are well behaved as $r \to \infty$.

In Sect.\ \ref{s: vector type QNF} we find that for $n$ even and for the vector type electromagnetic field, the radial function of the QNM takes the form (\ref{e: radial vector integer}). For $n=2$ this radial function behaves in the form 
\begin{align} \label{e: n=2 vector type}
 R_V &\approx (1-y) \,\, {}_2F_1(-p,\beta_V;\beta_V-\gamma_V-p+1;1-y) \nonumber \\
&\approx \frac{C_2}{r^2},
\end{align} 
as $r \to \infty$, where $C_2$ is a constant. Comparing the formulas (\ref{e: n=2 general asymptotic}) and (\ref{e: n=2 vector type}) we find that for $n=2$ we determine the QNF of the vector type electromagnetic field by canceling the leading term of the expression (\ref{e: n=2 general asymptotic}) ($C_1=0$), although both solutions are well behaved as $r \to \infty$. Motivated by this observation, for the previous puzzling cases of the scalar type electromagnetic field ($n=3,4,5$), it is convenient to study if we obtain well defined QNF when we cancel the leading term of the asymptotic behavior. 

From Eq.\ (\ref{e: radial scalar type}), in the limit $r \to \infty$,  we find that the function $R_S$ satisfies the differential equation
\begin{equation} \label{e: scalar type asymptotic}
 \frac{\dd^2 R_S}{\dd r^2} + \frac{3}{r} \frac{\dd R_S}{\dd r} - \frac{(n-2)(n-6)}{4} \frac{1}{r^2} R_S = 0,
\end{equation} 
whose solutions behave as (recall the formulas (\ref{e: radial vector asymptotic}) and (\ref{e: vector asymptotic behavior}))
\begin{equation}
 R_S \approx C_1 r^{|n-4|/2-1} + C_2 r^{-(|n-4|/2+1)}.
\end{equation} 
For $n \geq 7$ in the limit $r \to \infty$ the first term diverges, and the second goes to zero. Thus for these values of $n$ we can impose the boundary conditions a) and b) without a problem, but for $n=2,6,$ we find
\begin{equation} \label{e: scalar asymptotic n=2}
 R_S \approx C_1 + \frac{C_2}{r^2},
\end{equation} 
in a similar way to the vector type electromagnetic field for $n=2$ (see the expression (\ref{e: n=2 general asymptotic})). In the same limit, when $n=3,5$, we find that
\begin{equation} \label{e: scalar asymptotic n 3 5 black hole}
 R_S \approx \frac{C_1}{r^{1/2}} + \frac{C_2}{r^{3/2}},
\end{equation} 
and for $n=4$
\begin{equation}
 R_S \approx \frac{C_1}{r} + C_2 \frac{ln(r)}{r}.
\end{equation} 

Thus for $n=2,3,4,5,6,$ we see that both solutions are well behaved and for $n=3,4,5$, both solutions go to zero as $r \to \infty$. In the previous section for $n=2$ and $n=6$ we find well defined QNF for the scalar type electromagnetic field in the Lifshitz black hole (\ref{e: black hole Lifshitz}). Thus for these two values of $n$ we examine how the radial function of the QNM behaves as $r \to \infty$. In both cases this radial function is equal to (see the expression (\ref{e: scalar type integer even}))
\begin{equation}
 R_S = C_2 y^{-i \tilde{\omega}} (1-y) {}_2F_1(-p,\beta_S;\gamma_S;y) ,
\end{equation} 
and we notice that as $r \to \infty$ this radial function behaves in the form 
\begin{equation}
 R_S \approx \frac{C_2}{r^2} .
\end{equation} 
Comparing with the formula (\ref{e: scalar asymptotic n=2}), we notice that its leading term is fixed, as for the vector type electromagnetic field when $n=2$. Thus for the scalar type electromagnetic field  propagating in the $D$-dimensional Lifshitz black hole (\ref{e: black hole Lifshitz}) with $n=3,4,5,$ in what follows we determine the frequencies of the modes that are purely ingoing at the horizon and that as $r \to \infty $ its leading term is equal to zero.

From the expression (\ref{e: radial scalar boundary horizon}) for the radial function that satisfies the boundary condition of the QNM near the horizon, we get that for $n=3,5,$ the purely ingoing radial function takes the form 
\begin{equation}
  R_S = C_2  y^{- i \tilde{\omega}} (1-y)^{3/4} {}_{2}F_{1}(\alpha_S,\beta_S;\gamma_S;y) .
\end{equation} 
Using the Kummer formula (\ref{e: Kummer property y 1-y}), since $\gamma_S - \alpha_S - \beta_S$ is not an integer, we obtain that for $n=3,5$, in the limit $r \to \infty$ the radial function $R_S$ behaves as
\begin{align} \label{e: scalar type n=3,5 asymptotic}
 R_S \approx \frac{\Gamma(\gamma_S) \Gamma(\gamma_S-\alpha_S-\beta_S)}{\Gamma(\gamma_S-\alpha_S) \Gamma(\gamma_S - \beta_S)} \frac{r_+^{3/2}}{r^{3/2}} +  \frac{\Gamma(\gamma_S) \Gamma( \alpha_S + \beta_S - \gamma_S)}{\Gamma(\alpha_S) \Gamma(\beta_S)} \frac{r_+^{1/2}}{r^{1/2}} .
\end{align}
In this expression we see that the first term is subleading and the second term is leading. Hence to cancel the leading term we must impose the conditions (\ref{e: conditions scalar type}) and therefore when we cancel the leading term of the radial function for the scalar type electromagnetic field propagating in the Lifshitz black hole (\ref{e: black hole Lifshitz}) with $n=5$ we get the QNF (\ref{e: QNF scalar type}) and for $n=3$ the QNF are equal to
\begin{equation} \label{e: QNF scalar n=3}
 \omega = - i \frac{r_+^2}{l^3} \frac{1}{p+\tfrac{3}{4}} \left[ \left( p+\frac{1}{2} \right)(p+1) + \tilde{k}_S^2 \right],
\end{equation} 
(see the QNF (\ref{e: QNF scalar type n=2}) for the scalar type electromagnetic field when $n=2$).

In a similar way to the previous examples with $n$ even, for $n=4$ we know that the radial function satisfying the boundary conditions of the QNM near the horizon is
\begin{equation} \label{e: radial scalar type n=4}
  R_S = C_2  y^{- i \tilde{\omega}} (1-y)^{1/2} {}_{2}F_{1}(-p,\beta_S;\gamma_S;y) ,
\end{equation}
where we take $\alpha_S = -p$ as previously (see the expression (\ref{e: scalar type integer even})). Taking into account the property (\ref{e: hypergeometric polynomial}) of the hypergeometric function we deduce that in the limit $r \to \infty$ the radial function (\ref{e: radial scalar type n=4}) behaves as
\begin{equation}
 R_S \approx \frac{C}{r},
\end{equation} 
that is, in the expression (\ref{e: radial scalar type n=4}) the leading term is fixed. Therefore for $n=4$ the QNF of the scalar type electromagnetic field are determined by the expression (\ref{e: QNF scalar type}) with $n=4$, when we fix the leading part of the radial function. 

Thus if for the Lifshitz black hole (\ref{e: black hole Lifshitz}) with $n=3,4,5$, and the scalar type electromagnetic field we change the boundary condition imposed at the asymptotic region of the black hole, (that is, imposing that the leading term of the radial function is fixed, instead forcing to zero the radial function), then we obtain a well defined discrete spectrum of QNF. We believe that this result is more acceptable than our previous suggestion of a continuous QNF spectrum  for the  scalar type electromagnetic field.

It is known that for calculating the QNF of anti-de Sitter black holes have been imposed different boundary conditions at their asymptotic regions \cite{Berti:2009kk}, \cite{Konoplya:2011qq}, \cite{Breitenlohner:1982jf}--\cite{Dias:2013sdc}. Since for the scalar type electromagnetic field propagating in the Lifshitz black hole (\ref{e: black hole Lifshitz}) with $n=2,3,4,5,6,$ the two solutions of the radial equation are finite as $r \to \infty$, it is convenient to explore whether we can impose alternative boundary conditions in its asymptotic region. For example, for the scalar type electromagnetic field and  these values of $n$ we ask for the values of the frequencies for which the waves are purely ingoing near the horizon and  the subleading term is fixed as $r \to \infty$, that is, now the leading term is different from zero. For $n=3,5$, from the formula (\ref{e: scalar type n=3,5 asymptotic}), we see that to cancel the subleading term we must impose the condition
\begin{equation}
 \gamma_S - \alpha_S = - p, \qquad \textrm{or} \qquad \gamma_S - \beta_S = - p, \qquad  p=0,1,2,3,\dots,
\end{equation} 
from which we get that the frequencies are equal to 
\begin{equation} \label{e: frequencies leading}
 \omega_L = i \frac{r_+^2}{l^3} \left[ \frac{(n-2)^2/16 - \tilde{k}_S^2}{p+\tfrac{1}{4}} - \left(p + \frac{1}{4} \right) \right] . 
\end{equation} 
We notice that in the previous expression for $n=5$ and $\tilde{k}_S^2 < 1/2$ the fundamental mode ($p=0$) is unstable since for these values of the parameters the imaginary parts of the frequencies (\ref{e: frequencies leading}) satisfy $\im(\omega_L) > 0$, and we have a mode whose amplitude increases with time. Thus for some values of the parameters, when we fix the subleading term  of the radial function, we obtain unstable modes. Hence in contrast to anti-de Sitter black holes, to calculate the QNF for the Lifshitz black hole (\ref{e: black hole Lifshitz}) at its asymptotic region we can not impose the boundary condition that cancels the subleading term, but other boundary conditions (more complicated) can be used. Doubtless for Lifshitz black holes the calculation of the QNF for other boundary conditions at the asymptotic region is a problem that deserves additional study.

\section{Discussion}
\label{s: discussion}

In the $D$-dimensional asymptotically Lifshitz black hole (\ref{e: black hole Lifshitz}) with $n \geq 2$ for the vector type electromagnetic field we find the QNF (\ref{e: QNF vector type}). For the scalar type electromagnetic field moving in the black hole (\ref{e: black hole Lifshitz}) with $n \geq 4$ we obtain the QNF (\ref{e: QNF scalar type}), for $n=2$ the QNF (\ref{e: QNF scalar type n=2}), and the QNF (\ref{e: QNF scalar n=3}) for $n=3$. Notice that for $n=3,4,5,$ we must modify the boundary condition b) of the QNM in the asymptotic region, and fix the leading term in the expansion of the radial function as $r \to \infty$. From the expressions (\ref{e: QNF vector type}),  (\ref{e: QNF scalar type}),  (\ref{e: QNF scalar type n=2}), and (\ref{e: QNF scalar n=3}) for the QNF of the electromagnetic field in the Lifshitz black hole (\ref{e: black hole Lifshitz}) we find that the imaginary parts of the QNF satisfy $\im (\omega) < 0$, and since we choose a harmonic time dependence $\exp(-i \omega t)$, we assert that these QNM are stable because their amplitudes decay in time. We consider this result as a step towards establishing the classical stability of the Lifshitz black hole (\ref{e: black hole Lifshitz}) under small perturbations.

We also find that in the $D$-dimensional Lifshitz black hole (\ref{e: black hole Lifshitz}) with $n=3,4,5,$ the calculation of the QNF for the scalar type electromagnetic field requires a careful analysis. We find that for these values of $n$ this perturbation type has a discrete and stable spectrum of QNF if we impose a slightly different boundary condition at the asymptotic region. We show that fixing the leading term of the radial function as $r \to \infty$ we obtain well defined  QNF, in contrast to the continuous spectrum of QNF that we may get when the boundary condition b) is imposed. Nevertheless, see Ref.\ \cite{Estrada-Jimenez:2013lra} for a two-dimensional black hole with a continuous spectrum of QNF for the Klein-Gordon and Dirac fields.

Taking into account that the Hawking temperature for the asymptotically Lifshitz black hole (\ref{e: black hole Lifshitz}) is equal to \cite{Balasubramanian:2009rx}, \cite{Giacomini:2012hg}
\begin{equation}
 T_H = \frac{r_+^2}{2 \pi l^3},
\end{equation} 
we can write the QNF (\ref{e: QNF vector type}) for the vector type electromagnetic field in the form 
\begin{equation} \label{e: QNF vector type Hawking}
 \omega_V = -i \frac{ 2 \pi T_H}{p+\tfrac{1}{2} + \tfrac{n}{4}} \left[ (p+\tfrac{1}{2})^2 + \frac{n}{2} (p+\tfrac{1}{2}) + \frac{n-1}{4} + \tilde{k}_V^2\right],
\end{equation} 
and similarly for the QNF (\ref{e: QNF scalar type}), (\ref{e: QNF scalar type n=2}), and (\ref{e: QNF scalar n=3}) of the scalar type electromagnetic field. 

For the scalar type electromagnetic field propagating in the Lifshitz black hole (\ref{e: black hole Lifshitz}) with $n \geq 4$, from the QNF (\ref{e: QNF scalar type}) we find that the frequency of the fundamental ($p=0$) QNM takes the form
\begin{equation}
 \omega_F = - i \frac{r_+^ 2}{l^ 3}\frac{1}{n-2}\left(\frac{\hat{k}l}{r_+}\right)^2.
\end{equation} 
As we know, when the base manifold is $\mathbb{R}^n$ the eigenvalue $\hat{k}$ is continuous and satisfies $\hat{k}^2 > 0$. Thus, as $\hat{k} \to 0$, for the scalar type electromagnetic field we get that $|\im ( \omega_F) | \to 0$ and therefore we have a long lived fundamental QNM, that is, for this QNM its characteristic decay time $\tau = 1 / |\im (\omega_F)|$ is very large.

In contrast, for the scalar type electromagnetic field propagating in the Lifshitz black hole (\ref{e: black hole Lifshitz}) with $n=2$, from the QNF (\ref{e: QNF scalar type n=2}) we note that as $\hat{k} \to 0$ the frequency of the fundamental QNM goes to the value 
\begin{equation}
 \omega_F = - i \frac{4 r_+^2}{3 l^3},
\end{equation}  
and the characteristic decay time is equal to
\begin{equation}
 \tau = \frac{3 l^ 3}{4 r_+^2},
\end{equation} 
which is finite. Something similar happens with the QNF (\ref{e: QNF scalar n=3}) of the scalar type electromagnetic field in the Lifshitz black hole (\ref{e: black hole Lifshitz}) with $n=3$. For the QNM of the vector type electromagnetic field, in the limit $\hat{k}_V \to 0$, for its fundamental mode we also get that its characteristic decay time is finite.

We notice that in the $D$-dimensional Lifshitz black hole (\ref{e: black hole Lifshitz}) the QNF (\ref{e: QNF vector type}), (\ref{e: QNF scalar type}), (\ref{e: QNF scalar type n=2}), and (\ref{e: QNF scalar n=3}) of the electromagnetic field are purely imaginary. This fact indicates that in this  black hole, the electromagnetic QNM do not oscillate, they only decay, reminding us to a critically damped or overdamped classical system. This behavior is similar to that already found for some QNM, since at present time we know several spacetimes that possess purely imaginary QNF \cite{Giacomini:2012hg}, \cite{LopezOrtega:2009zx}--\cite{LopezOrtega:2007sr}, \cite{Estrada-Jimenez:2013lra}--\cite{LopezOrtega:2012vi}, and some of these black holes are asymptotically anti de Sitter, see for example Refs.\ \cite{Cardoso:2001bb}, \cite{Cardoso:2003cj}, \cite{Cordero:2012je}. Furthermore several black holes are poor oscillators in comparison with other natural systems, for example, for the Schwarzschild black hole its quality factor $Q = \tfrac{1}{2}|\re(\omega)/\im(\omega)|$ is approximately equal to the angular momentum number $K$, but for an atom $Q \approx 10^6$ \cite{Andersson}.

Recently Emparan and Tanabe \cite{Emparan:2014cia} show that in the limit $D \to \infty$ a large class of spherically symmetric black holes have a  universal set of QNM whose frequencies take the form 
\begin{equation} \label{e: Emparan QNF}
 \omega_D = \left[ \frac{D}{2} + K - \left(\frac{e^{i \pi}}{2} \left( \frac{D}{2}+ K \right) \right)^{1/3} a_p \right]\frac{1}{r_+},
\end{equation} 
where $r_+$ is the horizon radius, as before $K$ is the angular momentum number, and $-a_p$ are the zeroes of the Airy function \cite{Emparan:2014cia}. To compare with the results by Emparan and Tanabe we take the limit $D \to \infty$ ($n \to \infty$) of our expressions for the QNF of the electromagnetic field and of the formula (33) in Ref.\ \cite{Giacomini:2012hg} that gives the QNF of the Klein-Gordon field.

Thus from our expressions (\ref{e: QNF vector type}), (\ref{e: QNF scalar type}), and from the formula (33) of Ref.\ \cite{Giacomini:2012hg}, we get that in the limit $D \to \infty$ the QNF of the $D$-dimensional Lifshitz black hole (\ref{e: black hole Lifshitz}) simplify to 
\begin{align} \label{e: Lifshitz n infinity}
 \omega_V &= - i\frac{r_+^2}{l^3}(2p + 3), \nonumber \\
\omega_S &= - i\frac{r_+^2}{l^3}(2p ), \\
\omega_{KG} &= - i \frac{r_+^2}{l^3}(2p +2 ). \nonumber
\end{align}
Comparing the previous formulas with the expression (\ref{e: Emparan QNF}), we note some differences. Our expressions (\ref{e: Lifshitz n infinity}) are purely damped (purely imaginary), as our QNF (\ref{e: QNF vector type}), (\ref{e: QNF scalar type}), and the QNF (33) in Ref.\ \cite{Giacomini:2012hg}. Also they do not depend on the spacetime dimension $D$, in contrast to the result by Emparan and Tanabe \cite{Emparan:2014cia} whose frequencies are slowly damped and depend on the spacetime dimension $D$. Furthermore the limit $D \to \infty$ of the QNF of the $D$-dimensional Lifshitz black hole, that is the expressions (\ref{e: Lifshitz n infinity}) depend on the horizon radius as $r_+^2$, whereas the QNF (\ref{e: Emparan QNF}) show a dependence $1/r_+$.

We do not know the source of these differences between the expressions (\ref{e: Lifshitz n infinity}) for the QNF of the Lifshitz black hole (\ref{e: black hole Lifshitz}) and the suggested behavior by Emparan and Tanabe of the formula (\ref{e: Emparan QNF}), but we note that to calculate the QNF (\ref{e: Emparan QNF}) in Ref.\ \cite{Emparan:2014cia} is analyzed in detail the near horizon behavior of the field, and in this work to determine the QNF of the Lifshitz black hole (\ref{e: black hole Lifshitz}) it is relevant to consider the behavior of the field as $r \to \infty$ (as we showed in the previous sections).

\section{Acknowledgments}

This work was supported by CONACYT M\'exico, SNI M\'exico, EDI-IPN, COFAA-IPN, and Research Projects IPN SIP-20140832, IPN SIP-20131340, and IPN SIP-20131541.

\begin{appendix}

\section{Modes of the electromagnetic field in the Lifshitz spacetime}
\label{s: appendix}

To extend our previous results we determine the modes of the electromagnetic field in the Lifshitz spacetime (\ref{e: Lifshitz metric}). We  are looking for possible unstable solutions that satisfy the boundary conditions
\begin{enumerate}
 \item[i)] They are regular at $r=0$.
\item[ii)] They go to zero as $r \to \infty$.
\end{enumerate}

For the vector type electromagnetic field propagating in the Lifshitz spacetime (\ref{e: Lifshitz metric}), Eq.\ (\ref{e: vector type two-dimensional})  simplifies to the radial differential equation 
\begin{equation}
 \frac{\dd^2 R_V}{\dd r^2} + \frac{z+1}{r} \frac{\dd R_V}{\dd r} + \left( \frac{(\omega l^{z+1})^2}{r^{2z +2}} - \frac{n-2}{2}\left( 1+z+\frac{n-4}{2}\right)\frac{1}{r^2} - \frac{(\hat{k}_V l)^2}{r^4} \right) R_V = 0,
\end{equation} 
when we take $\Phi_V$ as in the formula (\ref{e: Phi vector}). As previously we take $z=2$. For this value of the critical exponent $z$ the previous differential equation transforms into 
\begin{equation} \label{e: vector type Lifshitz}
 \frac{\dd^2 R_V}{\dd r^2} + \frac{3}{r} \frac{\dd R_V}{\dd r} + \left( \frac{\check{\omega}^2}{r^{6}} - \frac{n^2-4}{4}\frac{1}{r^2} - \frac{\check{k}_V ^2}{r^4} \right) R_V = 0,
\end{equation} 
where we define the quantities $\check{\omega}$ and $\check{k}_V$ by $\check{\omega} = \omega l^3$ and $\check{k}_V=\hat{k}_V  l$.

For the scalar type electromagnetic field propagating in the Lifshitz spacetime (\ref{e: Lifshitz metric}), Eq.\ (\ref{e: scalar type two-dimensional})  simplifies to the ordinary differential equation 
\begin{equation} \label{e: scalar Lifshitz}
 \frac{\dd^2 R_S}{\dd r^2} + \frac{z+1}{r} \frac{\dd R_S}{\dd r} + \left( \frac{(\omega l^{z+1})^2}{r^{2z +2}} - \frac{n-2}{2}\left(\frac{n}{2}-1-z\right)\frac{1}{r^2} - \frac{(\hat{k} l)^2}{r^4} \right) R_S = 0,
\end{equation} 
when $\Phi_S = \exp (-i\omega t) R_S$ as in the expression (\ref{e: Phi scalar}). Taking $z=2$ we get that the previous differential equation becomes
\begin{equation} \label{e: scalar type Lifshitz}
 \frac{\dd^2 R_S}{\dd r^2} + \frac{3}{r} \frac{\dd R_S}{\dd r} + \left( \frac{\check{\omega}^2}{r^{6}} - \frac{(n-2)(n-6)}{4}\frac{1}{r^2} - \frac{\check{k}_S ^2}{r^4} \right) R_S = 0,
\end{equation} 
where $\check{k}_S = \hat{k} l$. Notice that for $n \geq 2$ the factor multiplying $1/ r^2$ in Eq.\ (\ref{e: vector type Lifshitz}) is greater or equal to zero, whereas in Eq.\ (\ref{e: scalar type Lifshitz}) it is negative for $n=3,4,5$.

To solve the ordinary differential equation (\ref{e: vector type Lifshitz}) we make the change of variable\footnote{We notice that this method also works for the differential equation (\ref{e: scalar type Lifshitz}) of the scalar type electromagnetic field.}
\begin{equation}
 u = \frac{1}{r},
\end{equation} 
to get that the function $R_V$ is a solution of the differential equation
\begin{equation}
 \frac{\dd^2 R_V}{\dd u^2} - \frac{1}{u} \frac{\dd R_V}{\dd u}  + \left( \check{\omega}^2 u^2 - \check{k}_V^2 - \frac{n^2-4}{4}\frac{1}{u^2} \right) R_V = 0.
\end{equation} 
Taking the radial function $R_V$ as
\begin{equation}
 R_V = \textrm{e}^{i\check{\omega}u^2/2} R_{V1} ,
\end{equation} 
and defining the new variable $v$ by
\begin{equation} \label{e: v definition}
 v = - i \check{\omega} u^2 ,
\end{equation} 
we find that the function $R_{V1}$ satisfies the differential equation 
\begin{equation}
 \frac{\dd^2 R_{V1}}{\dd v^2} - \frac{\dd R_{V1}}{\dd v} + \left(\frac{\check{k}_V^2}{4 i \check{\omega}} \frac{1}{v} - \frac{n^2 -4}{16} \frac{1}{v^2} \right)R_{V1} = 0 .
\end{equation} 

Proposing that the function $R_{V1}$ takes the form 
\begin{equation}
 R_{V1} = v^{A_V} R_{V2} ,
\end{equation} 
with the parameter $A_V$ being a solution to the equation
\begin{equation}
 A_V^2 - A_V - \frac{n^2-4}{16} = 0 ,
\end{equation} 
we obtain that the function $R_{V2}$ must be a solution of the ordinary differential equation 
\begin{equation}
 v \frac{\dd^2 R_{V2}}{\dd v^2} +(2 A_V - v) \frac{\dd R_{V2}}{\dd v} - \left( A_V + \frac{i \check{k}_V^2}{4 \check{\omega}} \right) R_{V2} = 0,
\end{equation} 
that is, of the confluent hypergeometric differential equation \cite{Guo-book}, \cite{NIST-book}
\begin{equation}
 v \frac{\dd^2 R_{V2} }{\dd v^2} + (b_V -v)\frac{\dd R_{V2} }{\dd v} - a_V R_{V2} = 0 ,
\end{equation} 
with the parameters $a_V$ and $b_V$ equal to
\begin{equation}
 a_V = A_V + \frac{i \check{k}_V^2}{4 \check{\omega}} , \qquad \quad b_V=2 A_V.
\end{equation} 
In what follows we choose
\begin{equation}
 A_V = \frac{1}{2} + \frac{n}{4},
\end{equation} 
that is,
\begin{equation} \label{e: Lifshitz vector parameters}
 a_V = \frac{1}{2} + \frac{n}{4} + \frac{i \check{k}^2_V}{4 \check{\omega}}  , \qquad \quad b_V= 1 + \frac{n}{2}.
\end{equation} 
We notice that the parameter $b_V$ of the previous formula is a positive half-integer for $n$ odd and a positive integer for $n$ even. 

From these facts, for the vector type electromagnetic field we find that the radial function takes the form \cite{Guo-book}, \cite{NIST-book}
\begin{equation}
 R_V = \textrm{e}^{-v/2} v^{b_V/2} \left\{C_1 U (a_V,b_V; v) + C_2  \textrm{e}^v U(b_V-a_V,b_V;\textrm{e}^{-i \pi} v) \right\}, 
\end{equation} 
where $C_1$, $C_2$ are constants and $U (a_V,b_V; v)$ is the Tricomi solution of the confluent hypergeometric differential equation. Since we choose a time dependence of the form $\exp(-i \omega t)$, and we are looking for unstable solutions, in what follows we assume that $\im (\check{\omega}) > 0$. From this assumption we obtain that the variable $v$ defined in the formula (\ref{e: v definition}) fulfills $\re(v) > 0$ and therefore the function satisfying the boundary condition i) at the origin ($v \to \infty$) is
\begin{equation} \label{e: solution Lifshitz vector}
 R_V = C_1  \textrm{e}^{-v/2} v^{b_V/2} U (a_V,b_V; v),
\end{equation}  
due to the another solution diverges as $r \to 0$.

First we take $n$ odd for the dimension of the base manifold. We point out that for $b$ different from an integer the Tricomi function $U (a,b; z)$ satisfies \cite{Guo-book}, \cite{NIST-book}
\begin{equation} \label{e: Tricomi confluent hypergeometric}
 U (a,b; z) = \frac{\Gamma(1-b)}{\Gamma(a-b+1)} {}_{1}F_{1}(a,b;z) + \frac{\Gamma(b-1)}{\Gamma(a)} z^{1-b} {}_{1}F_{1}(a-b+1,2-b;z),
\end{equation} 
where $ {}_{1}F_{1}(a,b;z)$ denotes the confluent hypergeometric function \cite{Guo-book}, \cite{NIST-book}, hence, as $r \to \infty$ ($v \to 0$) we find that the radial function (\ref{e: solution Lifshitz vector}) behaves as
\begin{equation} \label{e: vector infinity odd}
 R_V \approx   \frac{\Gamma(1-b_V)}{\Gamma(a_V-b_V+1)} v^{b_V/2}  +  \frac{\Gamma(b_V-1)}{\Gamma(a_V)} v^{1-b_V/2} .
\end{equation} 
Since we are considering $n$ odd ($n=3,5,7,\dots$), we see that $b_V \geq 5/2 $, thus, the first term of the expression (\ref{e: vector infinity odd}) goes to zero as $v \to 0$, whereas the second term diverges in this limit. Therefore to fulfill the boundary condition ii) we must impose the condition
\begin{equation} \label{e: vector condition n odd}
 a_V = -p, \qquad \quad p=0,1,2,\dots, 
\end{equation} 
from which we get the frequencies 
\begin{equation} \label{e: frequencies vector Lifshitz}
 \check{\omega}_V = -i \frac{\check{k}_V^2}{4}\frac{1}{p+\tfrac{n}{4}+\tfrac{1}{2}} .
\end{equation} 
We notice that for these frequencies $\im(\check{\omega}) < 0$, and hence we contradict our assumption $\im(\check{\omega}) > 0$. Thus we assert that in the Lifshitz spacetime (\ref{e: Lifshitz metric}) with $z=2$ and $n$ odd, for the vector type electromagnetic field  there is no unstable modes satisfying the boundary conditions i) and ii).

For $n$ even we can not use the property (\ref{e: Tricomi confluent hypergeometric}) since the quantity $b_V$ is a positive integer greater than 1, that is,  $b_V = 1 + N$ with $N=1,2,\dots$. Thus based on our previous results for $n$ odd, we take $a_V = -p$, with $p$ a nonnegative integer as in the formula (\ref{e: vector condition n odd}) and we search whether the radial function (\ref{e: solution Lifshitz vector}) fulfills the boundary condition ii) for this value of the parameter $a_V$. For $b_V$ a positive integer and $a_V$ a non-positive integer, the Tricomi function is equal to \cite{NIST-book} 
\begin{equation}\label{e: Tricomi both integers}
 U(-p,1+N;v) = (-1)^{-p} \sum_{s=0}^{p}  \binom{p}{s} (N + s + 1)_{p-s} (-v)^s .
\end{equation} 
From this expression we find that, as $v \to 0$, for $n$ even the radial function (\ref{e: solution Lifshitz vector}) behaves in the form 
\begin{equation}
 R_V \approx v^{b_V/2}.
\end{equation} 
Thus the radial function satisfies the boundary condition ii) and therefore for $n$ even we get again the frequencies (\ref{e: frequencies vector Lifshitz}) with $\im (\check{\omega}) < 0$. As for $n$ odd, this fact contradicts our assumption on the imaginary part of the frequencies. Hence in the Lifshitz spacetime (\ref{e: Lifshitz metric}) with $n$ integer, $n \geq 2$, for the vector type electromagnetic field we do not obtain unstable modes that fulfill the boundary conditions i) and ii).

In a similar way, for the scalar type electromagnetic field propagating in the Lifshitz spacetime (\ref{e: Lifshitz metric}) we find that the solutions of the radial differential equation (\ref{e: scalar type Lifshitz}) take the form \cite{Guo-book}, \cite{NIST-book}
\begin{equation}
 R_S = \textrm{e}^{-v/2} v^{b_S/2} \left\{C_1 U (a_S,b_S; v) + C_2  \textrm{e}^v U(b_S-a_S,b_S;\textrm{e}^{-i \pi} v) \right\}, 
\end{equation} 
with $C_1$, $C_2$ constants, and the parameters $a_S$, $b_S$ are equal to
\begin{equation} \label{e: Lifshitz scalar parameters}
 a_S = \frac{1}{2} + \frac{|n-4|}{4} + \frac{i \check{k}^2_S}{4 \check{\omega}}  , \qquad \quad b_S= 1 + \frac{|n-4|}{2} .
\end{equation} 
We see that the quantity $b_S$ is a positive half-integer for $n$ odd and a positive integer for $n$ even. Assuming that $\im(\check{\omega}) > 0$, as for the vector type electromagnetic field, we see that the regular solution as $v \to \infty$ is
\begin{equation} \label{e: solution Lifshitz scalar}
 R_S = C_1  \textrm{e}^{-v/2} v^{b_S/2} U (a_S,b_S; v) .
\end{equation}

When $n$ is odd ($b_S$ a half-integer) and considering the property (\ref{e: Tricomi confluent hypergeometric}) we obtain that for $v \to 0$ the previous radial function behaves as
\begin{equation} \label{e: scalar infinity odd}
 R_S \approx   \frac{\Gamma(1-b_S)}{\Gamma(a_S-b_S+1)} v^{b_S/2}  +  \frac{\Gamma(b_S-1)}{\Gamma(a_S)} v^{1-b_S/2} .
\end{equation} 
For $n=7,9,11,\dots$, we obtain that the previous formula takes the form
\begin{equation}
 R_S \approx   \frac{\Gamma(1-b_S)}{\Gamma(a_S-b_S+1)} v^{1/2 + |n-4|/4}  + \frac{\Gamma(b_S-1)}{\Gamma(a_S)} v^{1/2 - |n-4|/4} .
\end{equation} 
We note that the first term of the previous expression goes to zero as $v \to 0$ and the second term diverges in this limit. Hence to obtain that $R_S \to 0$ as $v \to 0$ we must impose the condition 
\begin{equation} \label{e: condition frequencies scalar type}
 a_S = -p, \qquad \quad p=0,1,2,\dots ,
\end{equation}  
from which we obtain the frequencies 
\begin{equation} \label{e: frequencies scalar}
 \check{\omega}_S = - \frac{i \check{k}_S^2}{4} \frac{1}{p + \frac{1}{2} +\frac{|n-4|}{4}} .
\end{equation} 
Owing to the frequencies (\ref{e: frequencies scalar}) fulfill $\im{\check{\omega}} < 0$, we contradict our assumption on the imaginary parts of the frequencies and in the Lifshitz spacetime (\ref{e: Lifshitz metric}) with $n$ odd, $n \geq 7$, we do not find unstable modes for the scalar type electromagnetic field.

From the expression for $b_S$ of the formulas (\ref{e: Lifshitz scalar parameters}), we find that for $n=3,5,$ the formula (\ref{e: scalar infinity odd}) for $R_S$ simplifies to 
\begin{equation} \label{e: scalar infinity 3-5}
 R_S \approx v^{1/4} \left\{ \frac{\Gamma(1-b_S)}{\Gamma(a_S-b_S+1)} v^{1/2}  +  \frac{\Gamma(b_S-1)}{\Gamma(a_S)}  \right\}.
\end{equation} 
For these two values of $n$, from the  previous expression we assert that for any value of the frequency $\check{\omega}$ the radial function fulfills $R_S \to 0$ as $v \to 0$, and hence it satisfies the boundary condition ii) without any additional condition on the frequencies $\check{\omega}$ when $\im(\check{\omega}) > 0$, suggesting the existence of unstable modes. Thus for the scalar type electromagnetic field propagating in the Lifshitz spacetime (\ref{e: Lifshitz metric}) with $n=3,5,$ we find that it is necessary a more careful analysis  of the boundary condition that we impose at the asymptotic region. 

As previously in Sect.\ \ref{s: scalar n 3 4 5}, for $n=3,5,$ here we study in detail the behavior of the radial function $R_S$ at the asymptotic region of the Lifshitz spacetime (\ref{e: Lifshitz metric}). We find that the differential equation (\ref{e: scalar type Lifshitz}) simplifies to Eq.\  (\ref{e: scalar type asymptotic}) in the limit $ r \to \infty $. Furthermore from the formula (\ref{e: scalar infinity 3-5}) we see that the radial $R_S$ behaves in the form 
\begin{equation} \label{e: radial behavior scalar n 3 5}
 R_S \approx  \frac{(-i \check{\omega})^{3/4} \Gamma(1-b_S)}{\Gamma(a_S-b_S+1)} \frac{1}{r^{3/2}}  +  \frac{(-i \check{\omega})^{1/4} \Gamma(b_S-1)}{\Gamma(a_S)}  \frac{1}{r^{1/2}} ,
\end{equation} 
in this limit. (Compare with the expected behavior given in the formula (\ref{e: scalar asymptotic n 3 5 black hole}).) That is, both solutions go to zero at the asymptotic region. In a similar way to the Lifshitz black hole (\ref{e: black hole Lifshitz}) to get an acceptable result we modify the boundary condition that we impose at the asymptotic region and we cancel the leading term of the expression (\ref{e: radial behavior scalar n 3 5}) taking $a_S = - p$ as in the formula (\ref{e: condition frequencies scalar type}). Therefore for $n=3,5,$ we get again the frequencies (\ref{e: frequencies scalar}) that fulfill $\im(\check{\omega_S}) < 0$. Hence we do not find unstable modes of the scalar type electromagnetic field when it propagates in the Lifshitz spacetime (\ref{e: Lifshitz metric}) with $n=3,5$, if we modify the boundary condition to be satisfied at the asymptotic region.

For $n$ even ($b_S$ a positive integer), from our previous results for $n$ odd, we take $a_S=-p$ as in the formula (\ref{e: condition frequencies scalar type}), and considering the property (\ref{e: Tricomi both integers}) of the Tricomi function we get that near $v=0$ the radial function $R_S$ of the formula (\ref{e: solution Lifshitz scalar}) behaves as
\begin{equation}
 R_S \approx v^{1/2 + |n-4|/4} ,
\end{equation} 
that goes to zero as $v \to 0$ for $n=2,4,6,\dots$. Thus for $n$ even we obtain the frequencies (\ref{e: frequencies scalar}) with $\im(\check{\omega}) < 0$, and therefore for the scalar type electromagnetic field propagating in the Lifshitz spacetime (\ref{e: Lifshitz metric}) with $z=2$ and $n$ even, we do not find unstable modes that fulfill the boundary conditions i) and ii).

In brief, for $n=3,5$ we do not obtain unstable modes for the scalar type electromagnetic field if the boundary condition that we impose at the asymptotic region is slightly different from ii), that is, in a similar way to the Lifshitz black hole (\ref{e: black hole Lifshitz}), as $r \to \infty$ we fix the leading term of the radial function to get well defined frequencies whose associated modes are stable. Thus for the vector and scalar type electromagnetic field propagating in the $D$-dimensional Lifshitz spacetime (\ref{e: Lifshitz metric}) we do not find unstable modes.

\end{appendix}

\end{document}